\documentstyle[11pt] {article}

\title{Preliminary Energy Considerations in a Relativistic Engine}

\author{Asher Yahalom$^a$ \\
$^a$ Ariel University, Kiryat Hamada POB 3, Ariel 40700, Israel\\
e-mails: asya@ariel.ac.il}

\begin{document}
\maketitle

\newcommand{\beq} {\begin{equation}}
\newcommand{\enq} {\end{equation}}
\newcommand{\ber} {\begin {eqnarray}}
\newcommand{\enr} {\end {eqnarray}}
\newcommand{\eq} {equation}
\newcommand{\eqs} {equations }
\newcommand{\mn}  {{\mu \nu}}
\newcommand{\sn}  {{\sigma \nu}}
\newcommand{\rhm}  {{\rho \mu}}
\newcommand{\sr}  {{\sigma \rho}}
\newcommand{\bh}  {{\bar h}}
\newcommand {\er}[1] {equation (\ref{#1}) }
\newcommand {\ern}[1] {equation (\ref{#1})}

\begin {abstract}
In a previous paper \cite{MTAY1} we have shown that Newton'n third law cannot strictly
hold in a distributed system of which the different parts are at a finite distance from each other.
This is due to the finite speed of signal propagation which cannot exceed the speed of light at vacuum,
which in turn means that when summing the total force in the system the force does not add up to zero.
This was demonstrated in a specific example of two current loops with time dependent currents, the
above analysis led to suggestion of a relativistic engine \cite{MTAY3,AY1}.
Since the system is effected by a total force for a finite period of time this means that the system acquires mechanical momentum and energy, the question then arises if we need to abandon the law of momentum and energy conservation. The subject of momentum conversation was discussed in
\cite{MTAY4}. Here some preliminary aspects of the exchange of energy between the mechanical part of the relativistic engine and the electromagnetic field are discussed. We also refer briefly to the material composition, structure and properties of metals that should be
used in a relativistic engine.
\vspace*{0.5 cm}
\vspace*{0.5 cm}
\end {abstract}

\vfill

\section{Introduction}

Special relativity is a theory of the structure of space-time.
It was introduced in Einstein's famous 1905 paper: "On the Electrodynamics of Moving Bodies" \cite{Einstein}.
This theory was a consequence of empiric observations and the laws of electromagnetism which were formulated in
the middle of the nineteenth century by Maxwell in his famous four partial
 differential equations \cite{Maxwell,Jackson,Feynman} which owe their current form to Oliver Heaviside \cite{Heaviside}.
One of the consequences of these equations is that an electromagnetic signal travels at the speed of light $c$,
which led people to believe that light is an electromagnetic wave. This was later used by Albert Einstein \cite{Einstein,Jackson,Feynman} to formulate his special theory of relativity which postulates that the speed of light in vacuum $c$ is the maximal allowed velocity in nature. According to the theory of relativity no object, message, signal (even if not electromagnetic) or field can travel faster than the speed of light in vacuum. Hence retardation, if someone at a distance $R$ from me changes something I may not know about it for at least a retardation time of $\frac{R}{c}$. This means that action and its reaction cannot be generated at the same time because of the signal finite propagation speed.

Newton's laws of motion are three physical laws that, together, laid the foundation for classical mechanics. They describe the relationship between a body and the forces acting upon it, and its motion in response to those forces. The three laws of motion were first compiled by Isaac Newton in his Philosophiae Naturalis Principia Mathematica (Mathematical Principles of Natural Philosophy), first published in 1687 \cite{Newton,Goldstein}. The laws are:

\noindent
First law:	When viewed in an inertial reference frame, an object either remains at rest or continues to move at a constant velocity, unless acted upon by a net force.

\noindent
Second law:	In an inertial reference frame, the vector sum of the forces F on an object is equal to the mass m of that object multiplied by the acceleration vector a of the object: $F = ma$.

\noindent
Third law:	When one body exerts a force on a second body, the second body \textbf{simultaneously} exerts a force equal in magnitude and opposite in direction on the first body.

According to the third law the total force in a system not affected by external forces is thus zero.
This law has numerous experimental verifications and seem to be one of the corner stones of physics. However, in light
of the previous discussion it is obvious that  action and its reaction cannot be generated at
the same time because of the finite speed of signal propagation, hence the
third law is false in an exact sense although it can be true for most practical application due to the high speed of signal propagation.
 Thus the total force cannot be null at a given time.

The locomotive systems of today are based on two material parts each obtaining momentum which is equal and opposite to the momentum
gained by the second part. A typical example of this type  of system is a rocket which sheds exhaust gas to propel itself. However, the above relativistic considerations suggest's a new type of motor in which the system is not composed of two material bodies but of a material body and field. Ignoring the field a naive observer will see the material body gaining
momentum created out of nothing, however, a knowledgeable observer will understand that the opposite amount of momentum is obtained
by the field as will be shown by exact calculation in this paper. Indeed Noether's theorem dictates that any system possessing
 translational symmetry will conserve momentum and the total physical system containing matter and field is indeed symmetrical under translations, while every sub-system (either matter or field) is not. This was already noticed by Feynman \cite{Feynman} (Feynman Lectures Vol. 2, 26-2 and 27-6). Feynman describes two orthogonally moving charges, apparently violating Newton's third law as the forces that the charges induce on each other do not cancel (last part of 26-2), this paradox is resolved in (27-6) in which it is shown that the momentum gained by the two charge system is balanced by the field momentum.

 In what follows we will assume that the magnetization and polarization of the medium are small and
therefore we neglect corrections to the Lorentz force suggested in \cite{Mansuripur}.

In a paper by Griffiths \&  Heald \cite{Griffiths} it was pointed out that strictly Coulomb's law and the Biot-Savart law determine
the electric and magnetic fields for static sources only. Time-dependent generalizations of these two laws introduced by Jefimenko
\cite{Jefimenko} were used  to explore the applicability of Coulomb and Biot-Savart outside the static domain.

In a previous paper we used Jefimenko's \cite{Jefimenko,Jackson} equation to discuss the force between two current carrying coils \cite{MTAY1}.
This was later expanded to include the interaction between a current carrying loop and a permanent magnet  \cite{MTAY3,AY1}.
Since the system is effected by a total force for a finite period of time this means that the system acquires mechanical momentum and energy, the question then arises if we need to abandon the law of momentum and energy conservation. The subject of momentum conversation was discussed in
\cite{MTAY4}. In this previous paper we discussed momentum conservation in a general system of given charge and current densities.
We have also calculated the force on the center of mass in the case of a particular system of two current carrying loops.
Finally we calculated the momentum gained by the material system and the momentum gained by the field and showed them to be equal
but with opposite directions.

In this paper some preliminary aspects of the exchange of energy between the mechanical part of the relativistic engine and the electromagnetic field are discussed. We also refer briefly to the material composition, structure and properties of metals that should be
used in a relativistic engine.

\section{Energy Conservation}

Any physical system with time translational symmetry must conserve energy according to Noether theorem.
In the case of a system with charge and current densities the energy conservation law takes the form \cite{Jackson}:
\beq
\frac{d E_{mech}}{dt}+\frac{d  E_{field}}{dt}= -\oint_S \vec S_p \cdot \hat n da.
\label{Econ}
\enq
In the above $E_{mech}$ is the mechanical energy of the system, $E_{field}$ the electromagnetic energy of the system
which is defined as:
\beq
E_{field}\equiv \int e_{field} d^3 x =\frac{\epsilon_0}{2} \int \left( \vec E^2 + c^2 \vec B^2 \right) d^3 x
\label{Efield}
\enq
and $\vec S_p$ is Poynting's vector defined as:
\beq
\vec S_p= \frac{1}{\mu_0} \vec E \times \vec B
\label{Poynting}
\enq

Energy and momentum conservation equations can be combined in a four dimensional relativistic formulation. In this formulation
one defines the four dimensional stress-energy tensor as follows:
\beq
\Theta^{\alpha \beta} = \left(
                                \begin{array}{cc}
                                  e_{field} & \frac{1}{c}\vec S_p \\
                                  \frac{1}{c}\vec S_p & -T_{i j} \\
                                \end{array}
                              \right)
\label{sten}
\enq
in which $T_{i j}$ is the Maxwell stress tensor \cite{Jackson}.
In terms of four dimensional stress-energy tensor it is possible to combine energy and momentum conservation in a single equation:
\beq
\int (\partial_\alpha \Theta^{\alpha \beta}+f^\beta) d^3 x = 0
\label{stencon}
\enq
In the above $\partial_\alpha$ is a partial derivative with respect to the four dimensional coordinates of
the relativistic formulation and:
\beq
\int f^\beta d^3 x =(\frac{d E_{mech}}{dt},\frac{d \vec P_{mech}}{dt})
\label{f}
\enq
$\vec P_{mech}$ is the total mechanical momentum of the system. However, despite the elegant relativistic formulation the content of the energy and momentum equations does not change. The relativistic formulation will not be discussed further in this paper.

\section{The case of two current loops }
\label{twocurlo}
\begin{figure}
\vspace{3cm} \includegraphics{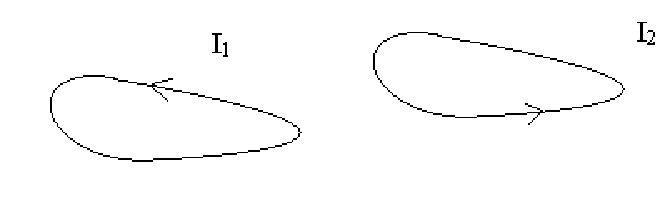}
\caption {Two current loops.}
 \label{twoloops}
\end{figure}

Consider two wires having segments of length $d\vec l_1,d\vec l_2$ located at $\vec x_1,\vec x_2$  respectively and
carrying currents $I_1, I_2$ (see figure \ref{twoloops}).

According to \cite{MTAY1} (equation (38)) the force on loop $_2$ generated by loop $_1$ takes the form:
\beq
\vec F_{21} = \frac{\mu_0}{4 \pi} I_2 (t) \sum_{n=0}^{\infty} \frac{I_1^{(n)}(t)}{n!}
(- \frac{1}{c})^n (1-n) \oint \oint R^{n-3} \vec R (d\vec l_2 \cdot d\vec l_1),
\label{F21ft3}
\enq
in which $\vec R \equiv \vec x_1 - \vec x_2$, $R \equiv |\vec R|$ and $\mu_0 =4 \pi 10^{-7}$ is the vacuum magnetic permeability.
We note that there is no first order contribution to the force. Hence the next contribution to
the force after the quasi-static term is second order.
Let us define the dimensionless geometrical factor $\vec K_{21n}$ as:
\beq
\vec K_{21n} = \frac{1}{h^n}  \oint \oint R^{n-3} \vec R (d\vec l_2 \cdot d\vec l_1) = -\vec K_{12n}.
\label{Kdef}
\enq
in the above $h$ is some characteristic distance between the coils. In terms of $\vec K_{21n}$ we can write \ern{F21ft3} as:
\beq
\vec F_{21} = \frac{\mu_0}{4 \pi} I_2 (t) \sum_{n=0}^{\infty} \frac{I_1^{(n)}(t)}{n!} (- \frac{h}{c})^n (1-n) \vec K_{21n}.
\label{F21ft4}
\enq
The force due to coil $2$ that acts on coil $1$ is:
\beq
\vec F_{12} = \frac{\mu_0}{4 \pi} I_1 (t) \sum_{n=0}^{\infty} \frac{I_2^{(n)}(t)}{n!} (- \frac{h}{c})^n (1-n) \vec K_{12n}.
\label{F12ft}
\enq
The total force on the system is thus:
\beq
\vec F_{T} = \vec F_{12} + \vec F_{21} =
\frac{\mu_0}{4 \pi}  \sum_{n=0}^{\infty} \frac{(1-n)}{n!} (- \frac{h}{c})^n  \vec K_{12n}
\left(I_1 (t) I_2^{(n)}(t) - I_2 (t) I_1^{(n)}(t)\right) .
\label{Ftotal2}
\enq
We note that the quasi-static term $n=0$ does not contribute to the sum nor does the $n=1$ term.
The fact that the retarded field "corrects" itself to first order in order to "mimic" a non retarded
field was already noticed by Feynman \cite{Feynman}.
Hence we can write:
\beq
\vec F_{T} = \frac{\mu_0}{4 \pi}  \sum_{n=2}^{\infty} \frac{(1-n)}{n!} (- \frac{h}{c})^n  \vec K_{12n}
\left(I_1 (t) I_2^{(n)}(t) - I_2 (t) I_1^{(n)}(t)\right) .
\label{Ftotal3}
\enq
We conclude that in general Newton's third law is not satisfied, taking the leading non-vanishing terms in the above sum
we obtain:
\ber
\vec F_{T} &\cong& -\frac{\mu_0}{8 \pi} (\frac{h}{c})^2  \vec K_{122}
\left(I_1 (t) I_2^{(2)}(t) - I_2 (t) I_1^{(2)}(t)\right) .
\label{Ftotal4}
\enr
Assuming that the total momentum of the system and the current derivatives are null at $t=0$, we obtain a mechanical linear momentum
as follows:
\beq
\vec P_{mech}=\int_0^t \vec F_{T} (t')dt' \cong -\frac{\mu_0}{8 \pi} (\frac{h}{c})^2  \vec K_{122}
\left(I_1 (t) I_2^{(1)}(t) - I_2 (t) I_1^{(1)}(t)\right) .
\label{Pmech1}
\enq
For simplicity we will now on assume a permanent current in loop $_2$ hence:
\beq
\vec P_{mech} \cong \frac{\mu_0}{8 \pi} I_1^{(1)}(t) I_2 (\frac{h}{c})^2  \vec K_{122}.
\label{Pmech1b}
\enq
For a calculation of $\vec K_{122}$ in particular geometries see \cite{MTAY1,MTAY3,AY1}. The kinetic mechanical energy associated
with this momentum is:
\beq
E_{mech}= \frac{ \vec P_{mech}^2}{2 M} = \frac{1}{2} \vec P_{mech} \cdot \vec v.
\label{Emech}
\enq
Were $M$ is the mass of the relativistic engine and $\vec v =  \frac{\vec P_{mech}}{M}$ is the engine velocity. This indicates that unlike the conservation of momentum \cite{MTAY4} which was independent of the mass and therefore of the velocity attained by the system, in the calculations of both the mechanical and electromagnetic energies the systems velocity and mass are of paramount importance.

\section{Field Energy}

Consider two sub-systems denoted system $_1$ and system $_2$ which are far apart such that their interaction is negligible.
In this case \er{Econ} is correct for each sub-system separately, that is:
\beq
\frac{d E_{mech~1}}{dt}+\frac{d  E_{field~1}}{dt}= -\oint_S \vec S_{p~1} \cdot \hat n da.
\label{Econ1}
\enq
\beq
\frac{d E_{mech~2}}{dt}+\frac{d  E_{field~2}}{dt}= -\oint_S \vec S_{p~2} \cdot \hat n da.
\label{Econ2}
\enq
Next we will put the two loops closer together such that they may interact but without modifying the
charge and the current densities of each of the subsystems. The total fields of the combined system are:
\beq
\vec E = \vec E_1 + \vec E_2, \qquad \vec B = \vec B_1 + \vec B_2
\label{Fields}
\enq
Since both the field energy \ern{Efield} and Poynting's vector \ern{Poynting}
are quadratic in the fields the following result is obtained:
\ber
E_{field} &\equiv& \frac{\epsilon_0}{2} \int \left( \vec E^2 + c^2 \vec B^2 \right) d^3 x =
E_{field~1}+E_{field~2}+E_{field~12}
\nonumber \\
E_{field~1} &\equiv& E_{E field~1}+E_{M field~1} \equiv \frac{\epsilon_0}{2} \int \left( \vec E_1^2 + c^2 \vec B_1^2 \right) d^3 x
\nonumber \\
E_{field~2} &\equiv& E_{E field~2}+E_{M field~2} \equiv \frac{\epsilon_0}{2} \int \left( \vec E_2^2 + c^2 \vec B_2^2 \right) d^3 x
\nonumber \\
E_{field~12} &\equiv& E_{E field~12}+E_{M field~12} \equiv \epsilon_0 \int \left(  \vec E_1 \cdot \vec E_2 + c^2 \vec B_1 \cdot \vec B_2 \right) d^3 x
\label{Efielddiv}
\enr
\ber
\vec S_p & \equiv & \frac{1}{\mu_0} \vec E \times \vec B  = \vec S_{p~1}+\vec S_{p~2}+ \vec S_{p~12}
 \nonumber \\
\vec S_{p~1} & \equiv & \frac{1}{\mu_0} \vec E_1 \times \vec B_1
 \nonumber \\
\vec S_{p~2} &\equiv & \frac{1}{\mu_0} \vec E_2 \times \vec B_2
 \nonumber \\
\vec S_{p~12} & \equiv & \frac{1}{\mu_0} \left( \vec E_1 \times \vec B_2 +\vec E_2 \times \vec B_1\right)
\label{Poyntingdiv}
\enr
Subtracting from \ern{Econ} the expressions given in \ern{Econ1} and \ern{Econ2}:
\ber
& &\hspace{-1cm}\frac{d (E_{mech}-E_{mech~1}-E_{mech~2})}{dt}+\frac{d  (E_{field}-E_{field~1}-E_{field~2})}{dt}
 \nonumber \\
&=& -\oint_S \left(\vec S_{p}-\vec S_{p~1}-\vec S_{p~2} \right) \cdot \hat n da.
\label{Econt-1}
\enr
taking into account \ern{Efielddiv} and \ern{Poyntingdiv} we arrive at:
\beq
\frac{d (E_{mech}-E_{mech~1}-E_{mech~2})}{dt}+\frac{d  E_{field~12}}{dt}
 = -\oint_S \vec S_{p~12} \cdot \hat n da.
\label{Econt0}
\enq
It is assumed that the electromagnetic loop currents do not self-propel and generate their own kinetic energy
due to the fields generated by their own sources, we also neglect at this stage heat dissipation from the system which is
included in $E_{mech}$ for any current loop of non zero resistivity. One can thus assume that the mechanical energy generated in each subsystem is negligible with respect to the mechanical energy generated in one subsystem due to the fields generated in the second subsystem and vice versa. Hence the self generated mechanical energies $E_{mech~1}$ and $E_{mech~2}$ are negligible and we obtain:
\beq
\frac{d E_{mech}}{dt}+\frac{d E_{field~12}}{dt}= -\oint_S \vec S_{p~12} \cdot \hat n da.
\label{Econt}
\enq

Let us now investigate the case that one system (say system $_2$) is static. In such a system their is no electric field and the magnetic field will fall as $\frac{1}{R^2}$ at distances $R$ far away from the system, this is different from the behaviour of fields in time dependent systems where
the field fall as $\frac{1}{R}$ \cite{Jackson,Jefimenko}. If we take the surface at infinity its area will grow as $R^2$, but
since $\vec S_{p~12}$ decrease as $\frac{1}{R^3}$  (as only system $_2$ is assumed to be static) in the limit of a large volume of integration the right hand surface integral vanishes and we obtain:

\beq
\frac{d  (E_{mech}+E_{field~12})}{dt}= 0
\label{Econt2}
\enq
Hence provided that there is no field or mechanical energy at $t=0$ we arrive at the result:
\beq
E_{mech}=-E_{field~12}.
\label{Econt3}
\enq
 We remark that if system $_2$ is magnetostatic (such as a constant current loop) then $\vec E_2 = 0$. In
such a cases:
\beq
E_{field~12} = E_{M field~12} = \frac{1}{\mu_0} \int \vec B_1 \cdot \vec B_2 d^3 x.
\label{Econt4}
\enq
It easy to see that this is just a mutual inductance energy term which has to do with the energetic price of putting
the too loop current close together from infinity. It exists also in the magnetostatic case (booth loops with constant currents) and has
nothing to do with a relativistic effect. It is of course not related to the mass or velocity of the relativistic engine. Hence under these conditions the {\bf kinetic} part of $E_{mech}$ is null and the relativistic engine effect is gone in contradiction to \ern{Emech}.

Notice, however, that in the above calculation we have neglected the change of the electromagnetic fields in the engine due
to its movement. As the engine sets into motion the electromagnetic field of the engine and each of its subsystems change. For
example system $_2$ which has no electric field in its own rest frame, but since this rest frame is in motion the magnetic field changes
in time in the laboratory frame and hence an electric field appears in the laboratory frame. It is exactly this change which causes
the term $E_{E field~12}$ to be different from zero.

\section{Field Energy of a System of two current loops}

We now return to the case of two current loops as described in section \ref{twocurlo}. However, now
we assume two types of time dependence. One is due to the intrinsic time dependence of the current flowing
through the loop and another is due to its movement as part of the relativistic engine. Thus the current density can be
written as:
\beq
\vec J(\vec x ,t) = \vec J' (\vec x - \vec x_c (t) ,t),
\label{Jp}
\enq
in which $\vec J' (\vec x ,t)$ is the current density in the moving frame of the relativistic engine and
$\frac{d \vec x_c (t)}{dt} = \vec v (t) $ is the velocity of that frame which is also the velocity of the relativistic engine.
The vector potential is given as \cite{Jackson}:
\beq
\vec A (\vec x,t) = \frac{\mu_0}{4 \pi} \int d^3 x' \frac{\vec J (\vec x',t_{ret})}{R}, \qquad \vec R \equiv \vec x-\vec x',
\qquad t_{ret} \equiv t-\frac{R}{c}.
\label{A}
\enq
This can be written as a power series in terms of $-\frac{R}{c}$ in the form:
\ber
\vec A (\vec x,t) &=& \frac{\mu_0}{4 \pi} \sum_{n=0}^{\infty} \frac{1}{n!} \int d^3 x' \frac{1}{R} (- \frac{R}{c})^n  \frac{d^n}{dt^n} \vec J (\vec x',t)
\nonumber \\
&=& \frac{\mu_0}{4 \pi} \sum_{n=0}^{\infty} \frac{1}{n!} \frac{d^n}{dt^n}  \int d^3 x' \frac{1}{R} (- \frac{R}{c})^n \vec J (\vec x',t)
\nonumber \\
&=& \frac{\mu_0}{4 \pi} \sum_{n=0}^{\infty} \frac{1}{n!} \frac{d^n}{dt^n}  \int d^3 x' \frac{1}{R} (- \frac{R}{c})^n  \vec J' (\vec x' - \vec x_c (t) ,t),
\label{A2}
\enr
Let us introduce a comoving integration variable: $\tilde{\vec{x}}=\vec x' - \vec x_c (t), \vec x'=\tilde{\vec{x}}+\vec x_c (t)$
such that: $R(t)=|\vec x' - \vec x|=|\tilde{\vec{x}}+\vec x_c (t)-\vec x|$ in such a comoving coordinate system we have:
\beq
\vec A (\vec x,t) = \frac{\mu_0}{4 \pi} \sum_{n=0}^{\infty} \frac{1}{n!} \frac{d^n}{dt^n}  \int d^3 \tilde{x}\frac{1}{R(t)} (- \frac{R(t)}{c})^n  \vec J' (\tilde{\vec{x}} ,t),
\label{A3}
\enq
For a thin and uniform current loop this can be written as:
\beq
\vec A (\vec x,t) = \frac{\mu_0}{4 \pi} \sum_{n=0}^{\infty} \frac{1}{n!} \frac{d^n}{dt^n} \left[ I (t) \oint d \tilde{\vec{l}} \frac{1}{R(t)} (- \frac{R(t)}{c})^n \right]  ,
\label{A4}
\enq
The zeroth order approximation takes the form:
\beq
\vec A (\vec x,t) \simeq \frac{\mu_0}{4 \pi} I (t)  \oint d \tilde{\vec{l}} \frac{1}{R(t)}.
\label{A5}
\enq
It easy to see that the first order correction to this vanishes as $\oint d \tilde{\vec{l}} = 0$.
We observe that \ern{A5} differs from the zeroth order approximation of equation (49) of \cite{MTAY4}
such that even if the current is constant in time the vector potential is not and thus an electric field
is generated.
The electric field $\vec E$ can be calculated as \cite{Jackson}:
\beq
\vec E = -\vec \nabla \Phi - \partial_t \vec A.
\label{E1}
\enq
In which $\Phi$ is the scalar potential. However, if we assume a Lorentz gauge and since there is no charge density in the current case the scalar potential $\Phi$ vanishes. Hence we obtain:
\beq
\vec E = - \partial_t \vec A = -\frac{\mu_0}{4 \pi} \left[\partial_t I (t)  \oint d \tilde{\vec{l}} \frac{1}{R(t)}+
I (t) \oint d \tilde{\vec{l}}\quad \frac{\vec v \cdot \vec R(t)}{R^3(t)} \right].
\label{E4}
\enq
It is well known that fields generated by a static source go as $\frac{1}{R^2}$ at infinity. But according to \ern{E4} this is also
true if the static source is moving. Hence for a moving relativistic engine, the leading contribution to the Poynting
flux of \ern{Econt} is null. Were the engine is composed of a constant current system $_2$ and a time dependent  current system $_1$.
This of course does not preclude system $_1$ from radiating. We also notice that $\vec v =  \frac{\vec P_{mech}}{M}=O(c^{-2})$, hence
the second term in \ern{E4} is usually smaller than the first. This rule has an exception in the case that the current is constant in time
and hence the first term is absent.

We now calculate the electric part of the electromagnetic energy of interaction:
\ber
 E_{E field~12}&=&\epsilon_0 \int \vec E_1 \cdot \vec E_2 d^3 x
 \nonumber \\
 &\hspace{-4cm} \simeq& \hspace{-2cm} \epsilon_0 \int   \left[-\frac{\mu_0}{4 \pi}\partial_t I_1 (t)  \oint_1 d \tilde{\vec{l_1}} \frac{1}{R_1(t)} \right] \cdot   \left[-\frac{\mu_0}{4 \pi} I_2  \oint_2 d \tilde{\vec{l_2}}\quad \frac{\vec v \cdot \vec R_2(t)}{R_2^3(t)} \right] d^3 x
\label{EN12}
\enr
in which we only considered the first order electric field terms in both the constant and time dependent current loops.
By reordering the terms this can be written as:
\beq
 E_{E field~12}= \frac{\mu_0}{(4 \pi c)^2} \partial_t I_1 (t)  I_2     \oint_1 \oint_2 d \tilde{\vec{l_1}} \cdot d \tilde{\vec{l_2}}
\vec v \cdot  \int \left[\frac{1}{R_1(t)} \frac{\vec R_2(t)}{R_2^3(t)} \right] d^3 x
\label{EN12b}
\enq
We shall now calculate the term:
\beq
\tilde{\vec{G}} = \int \left[\frac{1}{R_1(t)} \frac{\vec R_2(t)}{R_2^3(t)} \right] d^3 x
\label{Gt}
\enq
First let us introduce a change of variables:
\beq
\vec y = \vec R_2 = \vec x - \tilde{\vec x_2} - \vec x_c (t)
\label{y}
\enq
Since the integral $\tilde{\vec{G}}$ is calculated at a fixed point $\tilde{\vec x_2}$ it follows that $d^3 y =  d^3 x$ and:
\beq
\vec R_1 = \vec x - \tilde{\vec x_1} - \vec x_c (t)= \vec y + \tilde{\vec x_2} - \tilde{\vec x_1} \equiv \vec y - \vec R_{21}.
\label{R1}
\enq
This leads to the following expression for $\tilde{\vec{G}}$:
\beq
\tilde{\vec{G}} = \int y^{-3} \vec y \left|\vec y - \vec R_{21}\right|^{-1} d^3 y .
\label{G2}
\enq
This integral is now evaluated using a spherical coordinate system in which the "z" axis point
at the direction of $\vec R_{21}$. In this case $d^3 y = - y^2 dy d \cos \theta d \phi$ and $\tilde{\vec{G}}$
can be calculated as follows:
\beq
\tilde{\vec{G}} = -\int_{0}^{\infty}dy \int_{1}^{-1} d \cos \theta \int_{0}^{2 \pi} d \phi
 y^{-1} \vec y \left|\vec y - \vec R_{21}\right|^{-1} .
\label{G3}
\enq
Now:
\beq
\left|\vec y - \vec R_{21}\right| = \sqrt{y^2 + R_{21}^2 - 2 \vec y \cdot \vec R_{21}} = \sqrt{y^2 + R_{21}^2 - 2  y  R_{21} \cos \theta},
\label{yR}
\enq
In which we notice that the above expression is not dependent on the azimuthal angel $\phi$. Moreover, using
a cartesian set of unit vectors $\hat{y_1},\hat{y_2},\hat{y_3}$ one may write:
\beq
y^{-1} \vec y = \sin \theta \cos \phi \hat{y_1} +  \sin \theta \sin \phi \hat{y_2} + \cos \theta \hat{y_3},
\label{unity}
\enq
Thus it can easily be seen that there is component to $\tilde{\vec{G}}$ in the $\hat{y_1},\hat{y_2}$ directions as the
azimuthal integral vanishes. In the $\hat{y_3}$ direction the azimuthal integral is trivial and we obtain the result:
\beq
\tilde{\vec{G}} =2 \pi \hat{y_3} \int_{0}^{\infty}dy \int_{-1}^{1} d \cos \theta  \cos \theta
 \sqrt{y^2 + R_{21}^2 - 2  y  R_{21} \cos \theta}^{~-1} .
\label{G4}
\enq
Let us make a change of variables $s \equiv \cos \theta, y' \equiv \frac{y}{R_{21}}$ and notice that $\hat{y_3}= \hat{R}_{21}$ which
is a unit vector in the direction of $\vec R_{21}$, in terms of those variables we obtain a simpler representation of $\tilde{\vec{G}}$:
\beq
\tilde{\vec{G}} =2 \pi \hat{R}_{21} \int_{0}^{\infty}dy' \int_{-1}^{1} d s  s  \sqrt{y'^2 + 1 - 2  y' s}^{~-1} .
\label{G5}
\enq
However, we can evaluate analytically the $s$ integral to obtain:
\beq
\int_{-1}^{1} d s  s  \sqrt{y'^2 + 1 - 2  y' s}^{~-1} = \frac{2}{3} \left\{ \begin{array}{cc}
                                                                               \frac{1}{y'^2} & y' \geq 1 \\
                                                                               y' & y' < 1
                                                                             \end{array} \right. .
\label{sint}
\enq
And plugging this back into \ern{G5} we obtain:
\beq
\tilde{\vec{G}}  = 2 \pi \hat{R}_{21}.
\label{G6}
\enq
Having calculated $\tilde{\vec{G}}$ we are in a position to calculate $E_{E field~12}$ which now takes the form:
\beq
 E_{E field~12}= \frac{\mu_0}{8 \pi c^2} \partial_t I_1 (t)  I_2     \oint_1 \oint_2 (d \tilde{\vec{l_1}} \cdot d \tilde{\vec{l_2}})
(\vec v \cdot  \hat{R}_{21})
\label{EN12á}
\enq
We now take advantage of the definition given in \ern{Kdef} and notice that:
\beq
\oint \oint \hat{R} (d \vec l_1 \cdot  d \vec l_2) = \oint \oint R^{-1} \vec R (d \vec l_1 \cdot  d \vec l_2)
= - h^2 \vec K_{122}.
\label{Kdef2}
\enq
Thus we obtain:
\beq
 E_{E field~12} = - \frac{\mu_0}{8 \pi} I_2 \partial_t I_1 (t) \frac{h^2}{c^2} \vec v \cdot  \vec K_{122} = -\vec v \cdot \vec P_{mech} .
\label{Pf12g}
\enq
Comparing this to \ern{Emech} we observe that the change of the electrical energy due to the interaction of the two subsystems equals twice
the kinetic energy gained by the relativistic engine. This leads one to speculate that the other half goes into magnetic energy.
However, this calculation is beyond the scope of the current paper.

\section{Structure and Composition}

In order to avoid ohmic losses it is recommended to use super conducting materials for the construction of the current loops, however,
this may complicate the design due to the necessity to include cryogenic parts in the engine. The static part of the engine can be constructed
from a permanent magnet material which is a composite of rare earth elements such as Neodymium Ferron Boron or Samarium Cobalt, this will save a power supply but will probably increase the weight of the system. The use of standard copper wires is a viable alternative but in this case one will need to take into account also ohmic losses in the energy balance in addition to radiation losses and the energy taken by the relativistic engine.

\section{Conclusion}

We have shown in this paper that in general Newton's third law is not compatible with the principles of
special relativity and the total force on a two current loop system is not zero if one does not neglect retardation.
Still momentum and energy are conserved if one takes the field's momentum and energy into account and not just the mechanical momentum of the
material part of the device. This was shown here by direct computation.
The reader interested in the practical application of the device is referred to previous literature \cite{MTAY1,MTAY3,AY1}.


\begin {thebibliography} {99}

\bibitem {MTAY1}
Miron Tuval \& Asher Yahalom "Newton's Third Law in the Framework of Special ýRelativity" Eur. Phys. J. Plus (11 Nov 2014) 129: 240
 DOI: 10.1140/epjp/i2014-14240-x. (arXiv:1302.2537 [physics.gen-ýph]).ý
\bibitem {MTAY3}
Miron Tuval and Asher Yahalom "A Permanent Magnet Relativistic Engine" Proceedings of the Ninth International Conference on  Materials
Technologies and Modeling (MMT-2016) Ariel University, Ariel, Israel, July 25-29, 2016.
\bibitem {AY1}
Asher Yahalom "Retardation in Special Relativity and the Design of a Relativistic Motor". Acta Physica Polonica A, Vol. 131 (2017) No. 5, 1285-1288. DOI: 10.12693/APhysPolA.131.1285
\bibitem {MTAY4}
Miron Tuval and Asher Yahalom "Momentum Conservation in \\ a Relativistic Engine" Eur. Phys. J. Plus (2016) 131: 374. DOI: 10.1140/epjp/i2016-16374-1
\bibitem {Einstein}
A. Einstein, "On the Electrodynamics of Moving Bodies", Annalen der Physik 17 (10): 891–921, (1905).
\bibitem {Maxwell}
 J.C. Maxwell, "A dynamical theory of the electromagnetic field"
  Philosophical Transactions of the Royal Society of London 155: 459–512 (1865).
\bibitem {Jackson}
J. D. Jackson\index{Jackson J.D.}, Classical Electrodynamics\index{electrodynamics!classical}, Third Edition. Wiley: New York, (1999).
\bibitem {Feynman}
R. P. Feynman, R. B. Leighton \& M. L. Sands, Feynman Lectures on Physics, Basic Books; revised 50th anniversary edition (2011).
\bibitem {Heaviside}
O. Heaviside, "On the Electromagnetic Effects due to the Motion of Electrification through a Dielectric" Philosophical Magazine, (1889).
\bibitem {Newton}
I. Newton, Philosophiae Naturalis Principia Mathematica (1687).
\bibitem {Goldstein}
H. Goldstein , C. P. Poole Jr. \& J. L. Safko, Classical Mechanics, Pearson; 3 edition (2001).
\bibitem {Mansuripur}
M. Mansuripur, "Trouble with the Lorentz Law of Force: Incompatibility with Special Relativity and Momentum Conservation" PRL 108, 193901 (2012).
\bibitem {Griffiths}
D. J. Griffiths \&  M. A. Heald, "Time dependent generalizations of the Biot-Savart and Coulomb laws"
American Journal of Physics, 59, 111-117 (1991), DOI:http://dx.doi.org/10.1119/1.16589
\bibitem {Jefimenko}
Jefimenko, O. D., Electricity and Magnetism, Appleton-Century Crofts, New York (1966); 2nd edition, Electret Scientific, Star City, WV (1989).

\end{thebibliography}
\end{document}